\documentclass[10pt]{article}

\usepackage[latin1]{inputenc}
\usepackage[cyr]{aeguill}

\usepackage{amssymb}

\usepackage[T1]{fontenc}

\usepackage{graphicx}

\begin{document}

\begin{flushright}{CP3-06-04}\end{flushright}

\thispagestyle{empty}

\vspace*{15mm}

\begin{center}
{\LARGE   \textbf{The strong equivalence principle}}

\vspace*{2mm}

{\LARGE   \textbf{from a gravitational gauge structure}}
%
\end{center}

\vspace*{45mm}

 \begin{center}
{\large \textbf{J.-M. G\'erard}}\footnote{Email: gerard@fyma.ucl.ac.be}
\end{center}


\begin{center} \baselineskip0.4cm 
 Institut de Physique Th\'eorique and \\
 Centre for Particle Physics and Phenomenology (CP3), \\
 Universit\'e catholique  de Louvain,  \\
 B-1348 Louvain-la-Neuve, Belgium
\end{center}


\vspace*{35mm}

\begin{center}
{\large\textbf{Abstract}}

\vspace*{7mm}

\begin{minipage}{140mm}
\baselineskip0.4cm
  \hspace*{5mm}  Gravitational self-interactions are assumed to
 be determined by the covariant derivative acting on the Riemann-Christoffel
 field strength. Once imposed on a metric theory, this Yang-Mills gauge constraint
 extends the equality of gravitational mass and inertial mass to compact bodies
 with non-negligible gravitational binding energy. Applied to generalized
 Brans-Dicke theories, it singles out Einstein's tensor theory and Nordstr\"om scalar theory for
 gravity but also suggests a way to implement a minimal violation of the strong
 equivalence principle. 
\end{minipage}
\end{center}

\clearpage

\section*{Introduction}

The weak equivalence principle (WEP) rests upon the empirical equality
of the inertial and gravitational masses. It states that in an empty
space, namely a space with no matter present and no physical fields
except the (homogeneous) gravitational field, all test bodies fall
with the same acceleration. In practice, somebody in a free-falling
elevator would experience no apparent weight. From this `` happiest
thought '', Einstein inferred that all physical laws of special
relativity (electromagnetism included) remain valid in a sufficiently
small free-falling laboratory to eventually establish his succesful
general theory of relativity. In this theory, the gravitational
interactions of matter (and light) are characterized by a universal
coupling to the metric tensor $g_{\mu\nu}$. For the relativistic action of a
free particle, it simply amounts to substitute $g_{\mu\nu}(x)$ for the
Minkowski metric $\eta_{\mu\nu}$. 
Consequently any test particle propagates along
stationary paths and the track is always a geodesic of the curved
space-time, regardless of its mass. If the WEP is implemented without
reference to the Einstein non-linear field equations, what then does
privilege them ?  The careful observation of free-fall for compact
bodies (i.e bodies containing non-negligible gravitational
self-interactions, contrary to test bodies) should in principle
provide an answer to this question.

     The strong equivalence principle (SEP) naturally extends the
     universality of physical results to local gravitational
     experiments \cite{1}. It simply states that the free-fall of a compact
     body in empty space is also independent of its gravitational
     binding energy. Here, the ratio of internal gravitational binding
     energy to the total mass energy defines the `` compactness ''. From
     its typical value for human-size bodies, one easily concludes
     that present E\"otv\"os-like experiments only test the WEP, not
     the SEP. Yet, the Lunar Laser Ranging experiment beautifully
     corroborates the equivalence principle for celestial bodies : 
     the Moon orbit
     around the Earth does not appear to be polarized toward the
     Sun. In the first part of this paper, we raise
     this heuristic hypothesis about non-linear effects of gravity at
     the level of a fundamental principle.

    The WEP only rules the kinematics of point-like particles, not the
    dynamics of the gravitational fields. In fact, even after the
    Einstein field equations had been set up, it was thought that one
    had to demand in addition that the geodesic equation be the
    equation of motion of a test particle.  Eventually, it was
    realized that this can be deduced directly from the general
    covariant conservation of the energy-momentum tensor which is
    always valid in Einstein's theory.
 
     The SEP also governs the kinematics of extended bodies and
     appears to distinguish Einstein's theory from other relativistic
     theories of gravity. An effective violation of the SEP might be
     introduced through the explicit breakdown of energy-momentum
     conservation for compact bodies. In the second part of this
     paper, we suggest a covariant way to implement a minimal
     violation of the SEP without reference to the energy-momentum
     tensor for the matter fields.

    Both matter and self-interaction couplings are necessarily
    involved in the free-fall of bodies which carry a non-negligible
    amount of internal gravitational binding energy. Let us then first
    turn toward Yang-Mills theories where these couplings are precisely
    related.

\section*{Self-interaction Gauge Structure}

\subsection*{A. \ The case for strong interactions}

 Non-abelian gauge fields carry themselves the quantum numbers with which they
 interact. Indeed, in the classical equations of motion 
\begin{equation}
\partial_{\mu} F^{\mu\nu} = -g j^\nu - ig [A_{\mu},F^{\mu\nu}]
\end{equation}
 the first term on the rhs is the (gauge-invariant but not conserved) matter
 current, while the second one represents the non-linear self-coupling effect of
 the Yang-Mills fields \cite{2}. Consequently, these gauge fields act themselves
 as a source and their self-interaction  is  fixed by their  coupling to matter.
 In particular, the effective colour charges of the octet gluons (the gauge
 fields $A$ associated with strong interactions) and triplet quarks (the fermionic
 matter fields $F$) are equal up to well-defined SU(3) group theory factors. LEP
 precision measurements around the electroweak scale provide the corresponding
quadratic Casimir operators \cite{3}.
\[
\begin{array}{lllr}
\hspace{44mm}
 C_{A} &=& 2.89 \pm 0.01 \  (\textrm{stat.}) \pm 0.21 \  (\textrm{syst.}) &
\hspace{31mm}
(\textrm{2a})\\
\hspace{44mm}
 C_{F} &=& 1.30 \pm 0.01 \ (\textrm{stat.}) \pm 0.09  \ (\textrm{syst.}) &
(\textrm{2b})
\end{array}
\]
 which are in very good agreement with the non-abelian gauge structure of QCD
($C_{A} = 3$, $C_{F} = 4/3$) and rule out any abelian vector gluon model ($C_{A} =
0$, $C_{F} = 1/2$). They amount to a universal running of the three-gluon
 coupling defined by Eq.(1). This running as a function of energy points then at
 another genuine property of the strong interactions, namely quark \underline{and}
gluon confinement into hadronic bound states.

\setcounter{equation}{2}

 For pure gauge fields, i.e. without matter, Eq.(1) simply becomes the Yang-Mills
conditions
\begin{equation}
D_{\mu} F^{\mu\nu} = 0.
\end{equation}
The covariant derivative introduced in Eq.(3) defines then the field strength
\begin{equation}
[D_{\mu} , D_{\nu}] \equiv i F_{\mu\nu}
\end{equation}
 in a way similar to the definition of the curvature tensor for gravity \ldots

\subsection*{B. \ The case for gravitational interactions}

 If the gravitational field interacts with the matter fields through the general
 covariance which turns the ordinary derivative $\partial_{\mu}$ into the
covariant derivative $D_{\mu}$, any test body moves following geodesics,
regardless of its mass or internal structure. It results from this kinematics in
 curved space-time that the WEP is fulfilled. However, the gravitational field
 may also interact with itself, like non-abelian fields do. Indeed, the affine
connection $\Gamma^\lambda_{\mu\nu}$ corresponds to the gauge field $A_{\mu}$ and
the Riemann-Christoffel tensor $R^\sigma_{\ \lambda\mu\nu}$ corresponds to the
non-abelian field strength $F^{\mu\nu}$ with
\begin{equation}
[D_{\mu} , D_{\nu}]^\sigma_{\  \lambda} \equiv - R^\sigma_{\  \lambda\mu\nu}.
\end{equation}
 Compact bodies which differ by their binding energy could therefore fall with
 different accelerations if the gravitational self-interaction was not universal.
 But the present null results on an anomalous polarization of the Moon orbit
 around the Earth (the Nordtvedt effect \cite{4}) plead in favour of a single 
 three-graviton vertex known with a precision of 10$^{-3}$. Inspired by the
strong interaction gauge theory, we impose then the Yang-Mills conditions (3) on the
 curvature tensor :
 \begin{equation}
D_{\sigma} R^\sigma_{\  \lambda\mu\nu} = 0\,,
\end{equation}
in regions where matter is absent. Note that Eq.(6) does not require
the introduction of a metric but only defines a Riemannian manifold
for which the parallel-displacement field is sourceless. Thus
geodesics are not defined (yet) and the motion of a body cannot be
determined from Eq.(6) alone. However the resulting constraints
naturally embody the gravitational self-interaction in the cubic
$\Gamma\Gamma\Gamma$ terms. If the affine connection $\Gamma$
corresponds to the gauge potential $A$, gravitons gravitate the way
gluons glue. Consequently, a gravity theory in which Eq.(6) holds may
incorporate a suitable version of the SEP in the following sense : it
is an intrinsic property of gravitational dynamics, entirely
determined by the geometry of space-time. If the twenty constraints of
Eq.(6) turned out to over-determine the field equations of the theory, 
then this would imply a violation of the SEP. In other words,
Eq.(6) is not the full expression of the SEP but only a necessary
covariant condition to ensure its implementation. In particular, Eq.(6) does not
imply the WEP while the SEP does by definition. This latter point
becomes obvious once a metric is introduced.

 \bigskip
 
Let us express the covariant contraint (6) for the particular case of 
an asymptotically flat isotropic metric expanded in
the weak stationary field  $V(r) = -GM/r$:
\begin{eqnarray}
g_{00} &=& 1 + 2 \alpha \left({V\over c^2}\right) + 2 \beta \left({V\over
c^2}\right)^2 +
\mathcal{O} \left({1\over c^6} \right)\nonumber\\
g_{ij}  &=& - \delta_{ij}  \left \{1 - 2 \gamma \left({V\over c^2}\right) +
{3\over 2} \delta \left({V\over c^2} \right)^2 +
\mathcal{O} \left({1\over c^6} \right) \right\}.
\end{eqnarray}
The $\alpha, \beta, \gamma$ and $\delta$ dimensionless coefficients, normalized
to one for the Schwarzschild solution of general relativity, have to be
determined experimentally \cite{5}. From such a parametrization, we easily
obtain the following relations
\[
\begin{array}{lllr}
\hspace{22mm}
&\displaystyle  D_{\sigma} R^\sigma _{\  00n}   =   -\left({1\over 2c^4}\right)
(4\beta - \alpha\gamma -  3\alpha^2) \partial_{n} (\partial_{l} V  \partial^l V)
&\hspace*{7mm}
(\textrm{8a})\\
&&\\
& \displaystyle   D_{\sigma} R^\sigma _{\  \ell mn} = + \left({1\over c^4}\right)
(6\delta - 6\gamma^2 -
\alpha\gamma +
\alpha^2) (\partial_{l}\partial_{m} V \partial_{n}V - \partial_{l} \partial_{n}V
\partial_{m}V) 
 &\hspace*{7mm}  (\textrm{8b})
\end{array}
\]
if the ${\cal O} (1/c^6)$ terms are neglected. So, the third order
differential equations (6) for the metric $g_{\mu\nu}$ manifestly
contain unphysical solutions. For example, the exact solution with
$g_{00} = 1$ (i.e. $\alpha=\beta=0$, $\delta=\gamma^2$) possesses no
gravitational redshifts \cite{6} and violates thereby the WEP.
This nicely illustrates the fact that Eq.(6) is only a necessary
condition to fulfill the SEP.  Thus, at this level Eq.(6) should not
be regarded as the fundamental gravitational field equations derived
from some variational principle in a new theory of gravity \cite{7},
but rather as further tensorial constraints on any metric theory
incorporating by definition the WEP.

\section*{The SEP in Metric Theories}

Any metric theory of gravitation postulates the geodesic motion of test bodies such that the WEP corresponds to the zeroth-order condition $\alpha
\equiv 1$. The higher-order constraints imposed by Eq.(6) reduce then respectively to
\[
\begin{array}{lllr}
\hspace{50mm}
\eta  &\equiv& 4\beta - \gamma - 3 = 0   &
\hspace{43mm}
(\textrm{9a})\\
\hspace{50mm}
\eta' &\equiv& 6\delta - 6 \gamma^2 - \gamma + 1 = 0.   & (\textrm{9b})
\end{array}
\]
 Consequently, the Yang-Mills conditions (3) applied to the curvature field strength go
 beyond Einstein's heuristic hypothesis and imply that the non-linear parameters
 ($\beta$ \underline{and} $\delta$) are indepen\-dently determined by the
intrinsic space curvature ($\gamma$). The recent conjunction experiment with
Cassini spacecraft \cite{8} provides the range $\gamma - 1 = (2.1 \pm 2.3) \times
10^{-5}$. If the conditions (9) prove physical, we may already
 infer that  $\mid \beta-1\mid < 10^{-5}$  and $\mid \delta- 1\mid <  10^{-4}$ in
the vicinity of the Sun. These theoretical constraints are much stronger than
the existing observational bounds from the solar system \cite{9}.  At
present, the secular advance of Mercury's perihelion proportional to the
combination $(-\beta + 2\gamma +2)/3$ yields a weaker constraint on the
Eddington parameter $\beta$, $\mid \beta -1\mid < 3 \times 10^{-3}$. On the other
hand, the crucial parameter $\delta$ which appears only at the second order in
the light deflection angle
 (consistently expressed in terms of the physical impact parameter $b$) :

\setcounter{equation}{9}
\begin{equation}
\Delta = \left({4 GM \over c^2b}\right) \left\{{(1 +  \gamma) \over 2}  +  {(8 +
8\gamma  - 4\beta + 3\delta) \over 16}  {\pi GM \over c^2b} \right\}
\end{equation}
 is far from being constrained nowadays. Based on optical interferometry between
 two micro-spacecraft, the LATOR experiment \cite{10} aims at measuring
$\delta$ with a precision of 10$^{-3}$. But this experiment would simultaneously
reach the impressive level of accuracy of 1 to 10$^{8}$ for the parameter
$\gamma$.

\bigskip

 We are of course pleased to recover, as expected, the well-known Nordtvedt
 condition (9a) derived from a phenomenological relation \cite{11} between the
 gravitational mass $(m_{gr})$ and the inertial mass $(m_{in})$: 
\begin{equation}
{m_{gr} \over m_{in}} \approx  1 + \eta \left({\Omega \over mc^2}\right)
\end{equation}
 for compact bodies with non-negligible gravitational binding energy $\Omega $. Although
 the fraction of gravitational self-energy is only 4.5 $\times$ 10$^{-10}$ for
our planet, the Lunar Laser Range experiment confirms indeed that Earth and Moon
fall toward the Sun at equal rates with a precision of about 2 $\times$10$^{-13}$.
So, the gravitational binding energy equally contributes to the inertial mass and
to the gravitational mass with a precision given by \cite{9}
\begin{equation}
\mid \eta^{obs.} \mid = (4.4 \pm 4.5) \times10^{-4}. 
\end{equation}
 However the general covariance of Eq.(6) unavoidably requires a second condition
(9b) to be also fulfilled !

\bigskip

 To illustrate this important point, let us first briefly revive the Einstein-Grossmann
 ``Ent\-wurf''~\cite{12}. This `` outline ''    is
based on a spatially flat metric, $g_{ij}  = - \delta_{ij}$, and predicts an
advance of Mercury's perihelion of about 18" per century \cite{13}, i.e. 5/12 of
the observed value. In the weak field approximation, the corresponding Eddington
parameters are  $\gamma=\delta=0$ and $\beta = 3/4$,  respectively.   As a
consequence, the theory obeys the
Nordtvedt constraint (9a) but not (9b).  This can easily be understood from
the fact that the pure gravity kinetic term in the associated action
functional involves   ordinary derivatives of the metric field. Therefore,
general covariance is lost and the action is only invariant with respect to
arbitrary linear transformations.

\bigskip

 On the other hand, both constraints (9a) and (9b) are satisfied by the Einstein
 final theory \cite{14} with all Eddington parameters equal to unity $(\beta
\equiv 1, \gamma \equiv 1, \delta \equiv 1)$, but also
 by the Nordstr\"om scalar theory \cite{15}. The geometric reformulation \cite{16}
of this first consistent relativistic theory of gravitation leads indeed to a
special conformally flat metric,  $g_{\mu\nu} = (1+V/c^2)^2 \eta_{\mu\nu}$,
i.e. a finite set of Eddington parameters $(\beta = 1/2, \gamma = -1, \delta =
2/3)$. These particular values imply a retrogression of Mercury's perihelion of
1/6 of the observed magnitude as well as the vanishing of the deflection angle
$\Delta$ expressed in Eq.(10).

\bigskip

 The Bianchi identities,  $D_{\sigma} R^\sigma_{\  \lambda\mu\nu} + D_{\nu}
R^\sigma_{\ \lambda \sigma \mu} + D_{\mu} R^\sigma_{\ \lambda \nu \sigma}=0$, 
allow us to extend the analysis of the Yang-Mills conditions beyond the weak
field approximation. Indeed, these identities imply that the basic Eq.(6) is
equivalent to 
\begin{equation}
D_\nu R_{\lambda \mu} - D_{\mu} R_{\lambda \nu} = 0.
\end{equation}
 A contraction of Eq.(13) yields the necessary condition  $D_\nu R^\nu_{\ \mu} =
\partial_{\mu} R$. A direct
 comparison with the contracted Bianchi identities, 2  $D_{\nu} R^\nu_{\ \mu} =  
\partial_{\mu}R$, implies therefore a constant scalar curvature $R$. But our
hypothesis of an asymptotically flat metric eventually requires  $R$ to be
vanishing. Using then the standard decomposition of the Riemann tensor
$R^\sigma_{\ \lambda\mu\nu}$ into the Weyl tensor $W^\sigma_{\  \lambda\mu\nu}$
and the Ricci tensor  $R_{\lambda\mu}$, one easily infers that Eq.(13) is, in its
turn, equivalent to
\[
\begin{array}{lllr}
\hspace{50mm}
&& D_{\sigma}W^\sigma_{\ \lambda\mu\nu}= 0 &
\hspace{59mm}
(\textrm{14a})\\
\hspace{50mm}
&& R=0.      & (\textrm{14b})
\end{array}
\]
We immediately conclude from Eqs.(13) and (14) that both Einstein tensor
theory $(R_{\lambda\mu}= 0)$ and Nordstr\"om-Einstein-Fokker scalar theory
$(W^\sigma_{\ \lambda\mu\nu}= 0, R = 0$ are in fact exact vacuum solutions of Eq.(6).

\bigskip

 At the 1/c$^4$ order in the weak field approximation, the tensorial Eqs.(14a)
and (14b) respectively constrain two linear combinations of the $\eta$ and $\eta'$
parameters: 
\[
\begin{array}{lllr}
\hspace{51mm}
 (2\eta - \eta') &=&0 &
\hspace{60.5mm}
(\textrm{15a})\\
\hspace{51mm}
(\eta + \eta') &=&0.    & (\textrm{15b})
\end{array}
\]
 Consequently, $\eta'$ has definitely to be on an equal footing with the Nordtvedt
parameter
 $\eta$ defined in Eq.(11). This nicely confirms our intuition that Eq.(6)
provides a covariant formalism for the SEP in the general framework of
metric theories. Such a formalism seems to single out one pure tensor and one
pure scalar theory of gravity.  It is therefore worth  checking if they both
 survive within a restricted class of metric theories which naturally emerge as 
low-energy approximations of superstring or Kaluza-Klein theories.

\section*{The SEP in Tensor-Scalar Theories}

\setcounter{equation}{15}

Let us consider the following tensor-scalar $(TS)$ action
 \begin{equation}
S_{\textrm{\tiny\textit{TS}}} (\omega) = - \left({c^4\over 16\pi}\right) \int
d^4x
\sqrt{-g}
\left\{\phi R - {\omega  (\phi)\over \phi} g^{\mu\nu} \partial_{\mu} \phi 
\partial_{\nu}\phi\right\} +
S_{\scriptsize \textrm{Matter}} (g_{\mu\nu}, \psi)
\end{equation}
involving in principle an arbitrary function $\omega (\phi)$ of the scalar field
$\phi$.

\subsection*{A. \ Constant-$\omega$}

 If $\omega(\phi)$ is a constant parameter $\omega_{0}$, we recover the original
Brans-Dicke $(BD)$ theory \cite{17}. In this theory, Eq.(11) simply amounts to a
spatial variation of the Newton constant :
\begin{equation}
G_{\scriptsize\textrm{lab}} (r) \approx G_{\infty}  \left\{1 -
\eta_{\textrm{\tiny\textit{BD}}}
{V(r)\over c^2}\right\}
\end{equation}
 where $G_{\infty}$  is the value of $G$ measured far from the gravity source. Such
 a variation of the gravitational coupling obviously violates the universality of
 free-fall for compact bodies, i.e. the SEP. The inner structure of a compact
 body is indeed sensitive to changes of the gravitational `` constant ''.

 \bigskip

 In the weak field approximation, we easily obtain
  \begin{equation}
  \begin{array}{lll}
  \eta_{\textrm{\tiny\textit{BD}}} &=& 1-\gamma \nonumber\\
  \eta'_{\textrm{\tiny\textit{BD}}} &=& 2\gamma (\gamma-1)  
  \end{array}
\end{equation}
 with  $\gamma = (\omega_{0}+1)/(\omega_{0}+2)$. Consequently, Eqs.(9) are only
fulfilled by general relativity $(\omega_{0}=\infty)$. This result can easily be
extended beyond the weak field approximation. Indeed, variations of the BD
action with respect to the gravitational fields imply that the scalar $\phi$
plays the role of a source for the Riemann-Christoffel tensor :
\begin{equation}
D_{\sigma} (\phi R^\sigma_{\ \lambda\mu\nu}) = - (\omega_{0} +1) (R_{\lambda
\mu} \phi_{\nu} - R_{\lambda \nu} \phi_{\mu}).
\end{equation}
The basic Yang-Mills condition (6) requires  a constant
scalar field and the tensor-scalar action (16) reduces to the
Hilbert one for $G\phi = 1$. Note that the rhs of Eq.(19) vanishes for
$\omega_{0} = -1$, a value predicted by duality in the graviton-dilaton
low-energy effective superstring action \cite{18}. Such is not the case for the
low-energy limit of a (4 + $\varepsilon$) dimensional Kaluza-Klein theory
characterized by  $\omega_{0} + 1 = 1/\varepsilon$.

\subsection*{B. \  Variable-$\omega$}

 We have seen that the SEP requires 
$\gamma_{\textrm{\tiny\textit{BD}}}$ = +1.
 This is not a surprise since the Brans-Dicke theory obeys
$\beta_{\textrm{\tiny\textit{BD}}}$ = 1, i.e. rules
out the Nordstr\"om-Einstein-Fokker theory ($\beta$ = 1/2) from the start. But if
$\omega(\phi)$ is an arbitrary function of the scalar field, $\beta$ is now 
proportional to its  first derivative $\omega'(\phi)$. Remarkably, we still have
an exact relation between the parameters $\beta,\gamma$ \underline{and} $\delta$
:
\begin{equation}
 \delta_{\textrm{\tiny\textit{TS}}}  = {4\over 3}
(\beta_{\textrm{\tiny\textit{TS}}} -1) + {1\over 6} 
(8\gamma_{\textrm{\tiny\textit{TS}}}^2
-\gamma_{\textrm{\tiny\textit{TS}}} -1)
\end{equation}
 such that one linear combination of $\eta$ and $\eta'$ can be directly
expressed in terms of the space curvature parameter:
\begin{equation}
2\eta_{\textrm{\tiny\textit{TS}}} - \eta'_{\textrm{\tiny\textit{TS}}} = 2
(1-\gamma^2_{\textrm{\tiny\textit{TS}}})
\end{equation}
 even if $\omega(\phi)$ is not known! Consequently, a necessary condition to
 preserve the SEP in any tensor-scalar theory is
$\gamma_{\textrm{\tiny\textit{TS}}} = \pm 1$. This
generic result on the SEP is in agreement with \cite{19}. Combined with the
 constraint (9a) and the relation (20), it singles out both Einstein and
Nordstr\"om-Einstein-Fokker metrics at the \underline{full} $1/c^4$ order in the weak
field approximation. 
The general static spherically symmetric solution of Eq.(6), given as
an inverse-power-series in \cite{19a}, proves that $\alpha=\gamma=+1$
is sufficient to settle the Schwarzschild solution of general relativity.
Yet, our covariant formalism (14) allows us to go again beyond such a
result.

\bigskip

 The linear combination of $\eta$ and $\eta'$ in Eq.(21) is precisely associated
with the tensor $D_{\sigma} W^\sigma_{\ \lambda\mu\nu}$ in the weak field
approximation (see Eq.(15a)). Following then  Eq.(14), we still have the
condition of vanishing scalar curvature at our disposal :
\begin{equation}
R = \left\{\omega-3 {\omega' \phi \over (2\omega +3)}\right\} {\phi^\alpha
\phi_{\alpha}
\over \phi^2}=0.
\end{equation}
 This second condition completely fixes the arbitrary functional $\omega(\phi)$
appearing in the tensor-scalar action (16):
\begin{equation}
\omega(\phi) = - \left({3\over 2}\right) {G\phi \over (G\phi-1)}.
\end{equation}
A simple scalar field redefinition:
\begin{equation}
G\phi = 1 - {\kappa \varphi^2\over 6} >0
\end{equation}
  with $\kappa$ = $8\pi G/c^4$ leads then immediately to the improved   
 Einstein-massless scalar theory advocated by Deser \cite{20}. The resulting
action
\begin{equation}
S_{R=0} = - \left({1\over 2\kappa}\right) \int d^4x \sqrt{-g} R + \int d^4x
\sqrt{-g} \left\{\left({\varphi^2\over 12}\right) R + {1\over 2} g^{\mu\nu} 
\partial_{\mu} \varphi \partial_{\nu} \varphi\right\}
\end{equation}
 corresponds indeed to gravity conformally coupled to a scalar field
 \cite{21}.
But we have shown that it may as well be considered as the generalized
Brans-Dicke tensor-scalar theory with a vanishing scalar curvature
$R$. From this latter point of view, the other condition (14a) is also
satisfied if either the scalar (helicity-0) or the tensor (helicity-2)
degree of freedom is frozen ($\varphi \to 0$ or $g_{\mu\nu} \to
\eta_{\mu\nu}$, respectively). In the first case, the action simply
reduces
 to the Hilbert one and we are in the presence of the Einstein tensor
 theory. In
the second case, a massless scalar freely propagates in the Minkowski
space-time and
 we end up with the original Nordstr\"om scalar theory \cite{15}.

\bigskip

  We conclude that only the Einstein and Nordstr\"om-Einstein-Fokker metric
 theories do comply with the SEP as defined by Eq.(6). From an observational
viewpoint, it is  obvious that perihelion advance and light deflection
measurements ($\Delta \neq$ 0) exclude the latter in favour of the former.
However, it would be more satisfactory if one could already discriminate them at
the theoretical level.

\section*{Toward a Theory with Minimal Violation of the SEP}

The sub-class of tensor-scalar theories defined by:
\begin{equation}
2\omega(\phi) + 3 = {a\over (1-G\phi)}
\end{equation}
 with $a\neq 0$, is rather interesting since $\eta$ and $\eta'$ are then
separately expressed in terms of the crucial Eddington parameter $\gamma$:
\begin{equation}
\begin{array}{lll}
&\displaystyle \eta_{a} = (1-\gamma^2) {(1+a) \over 2a}\\
&\\
& \displaystyle \eta'_{a} = 2 (1-\gamma^2) {(1-a)\over 2a}. 
\end{array}
\end{equation}
For $\gamma \neq \pm 1$, the Yang-Mills conditions (3) imposed on the curvature
tensor are not fulfilled and the value of the parameter $a$ characterizes any
violation of the SEP (see Fig.1).

\bigskip

 The case $a = -1$ corresponds to the well-known `` constant-$G$ ''  theory of
Barker \cite{22}. Consequently, this theory violates the SEP  since
$\eta$
= 0 but  $\eta' \neq 0$.  Note that the same holds true for the complementary
case $a = +1$ (i.e., $\eta'$ = 0  but $\eta \neq 0$).

\bigskip

 For $a = + 3$, we recover the `` zero-$R$ '' theory defined in
 Eq.(23), with
$\eta + \eta'=0$ in the weak field approximation.We have seen from
Eq.(25) that this covariant theory provides a smooth interpolation
between the pure tensor and the pure scalar theories respecting the
SEP (see the curve $R=0$ in Fig.1). A minimal violation
 of the SEP is thus expected here.

\begin{figure}[h!]
\hspace*{-5mm}
\includegraphics[width=18cm,keepaspectratio]{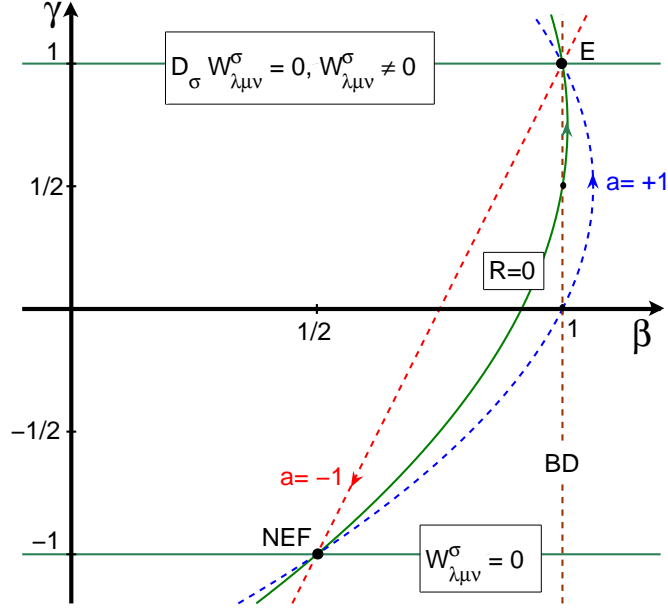}
\begin{center}
\begin{minipage}{100mm}
\caption{\footnotesize Tensor-scalar theories versus the Strong Equivalence
Principle. \qquad  Only the Einstein (E)
and Nordstr\"om-Einstein-Fokker (NEF) theories comply with the gauge conditions
(14). The dashed vertical line corresponds to the Brans-Dicke (BD) models
while the dashed curves represent the Barker theory $(a=-1)$ and its
complementary $(a=+1)$ defined by Eq.(26). Arrows indicate possible
attractors within FLRW cosmology.}
\end{minipage}
\end{center}
\end{figure}

\bigskip

 In fact, an attractor mechanism toward $G\phi = 1$ for  $a \gg 1$ in Eq.(26)
 has already been discussed in the Friedmann-Lemaitre-Robertson-Walker (FLRW)
 cosmology (i.e., with a time-dependent evolution for the scalar auxiliary field)
 \cite{23}. This mechanism is most easily described in the so-called Einstein
frame where the scalar and not the ordinary particle follows the geodesic
determined by the metric. The corresponding action
\begin{equation}
S_{E} = - \left({1\over 2\kappa}\right) \int d^4x \sqrt{-g} \{R-2g^{\mu\nu}
 \partial_{\mu} \sigma \partial_{\nu} \sigma\}
 + S_{\scriptsize \textrm{Matter}} \{A^2 (\sigma) g_{\mu\nu}, \psi\}
 \end{equation}
 is obtained through a suitable conformal transformation on the
metric originally defined in the Jordan frame (see Eq.16):
\begin{equation}
g_{\mu\nu} \to A^2 (\sigma) g_{\mu\nu}
\end{equation}
 with $G\phi = A^{-2} (\sigma)$ and $2\omega (\phi) + 3 = (A'/A)^{-2}$. In the
Einstein frame (28), the $\sigma$ field's cosmological evolution is just
analogous \cite{24} to the damped motion of a particle in the potential $\ln
A(\sigma)$.

 \bigskip
 
 For the sub-class of tensor-scalar theories defined by Eq.(26), we obtain
the corresponding conformal factor
\begin{equation}
A(\sigma) = \cosh   \left({\sigma \over \sqrt{a}}\right),
\end{equation}
such that general relativity $(\sigma=0)$ is the only point of equilibrium. If
 $a < 0$, a singular relaxation toward  $A = 0$  is generic. Yet, for $a > 0$,
 $A(\sigma) \geq 1$ and only a smooth relaxation toward  $A = 1$, i.e. general
relativity, is possible.  The `` constant-$G$ ''  theory $(a = -1)$  exhibits indeed a
singular attractor toward pure scalar gravity $(A \to 0)$. On the other, the `` zero-$R$ '' theory  $(a = +3)$  with positive $\omega(\phi)$ turns out to
contain a natural attractor toward general relativity $(A\to 1)$. Consequently,
cosmological dynamics provides  a way to disentangle the two theories
respecting the SEP in the static approximation (see arrows in  Fig.1). 

Let us therefore introduce matter ($\psi$), radiation ($\gamma$) as well as a cosmological constant ($\Lambda$) in the action $S_{R=0}$ defined by Eq.(25). The resulting field equations read then:
\begin{eqnarray}
&&R_{\mu\nu} -\frac12 R g_{\mu\nu} = \kappa ( 
T^\varphi_{\mu\nu}+T^\psi_{\mu\nu}+
T^\gamma_{\mu\nu}+T^\Lambda_{\mu\nu})\nonumber\\
&&D^\mu D_\mu \varphi = \frac{R}{6} \varphi\,. 
\end{eqnarray}
The massless field $\varphi$ behaves like a radiation field (i.e., 
$T^\varphi=T^\gamma=0$). So, we immediately obtain the equation
\begin{equation}
R= -\kappa ( T^\psi + T^\Lambda)\,.
\end{equation}
In the homogeneous FL cosmology, the RW metric is conformal to the Minkowski one (i.e., $W^\sigma_{~\lambda\mu\nu}=0$) 
such that Eq.(6) simply becomes
 \begin{equation}
D_{\sigma} R^\sigma_{\  \lambda\mu\nu} =  
\frac{\kappa}{6} \left( g_{\lambda \nu} \partial_\mu - g_{\lambda \mu} \partial_\nu \right) T^\psi
\end{equation}
when matter is present. Note that the scalar field ($\varphi$), 
the radiation ($\gamma$) and the cosmological constant ($\Lambda$) do not modify 
the vacuum contraint (6) in this theory.

For an initial value of the field $\varphi$ smaller than the
Planck mass (i.e., $\varphi_{\rm initial} \lesssim \kappa^{-\frac12})$, in agreement
with Eq.(24), the time-dependent solution of Eqs.(31) implies an upper
bound on the present space curvature, $(\gamma-1)_{\rm today} \lesssim
10^{-4}$. Current observations in the vicinity of the Sun already put
a more severe constraint on $\gamma$. The LATOR experiment could in principle 
probe the theory defined by Eq.(25) over four orders of magnitude.

\section*{Conclusion}

 The weak equivalence principle (WEP) settles the kinematics of
 test particles (space-time tells mass how to move), but not the 
 dynamics of gravity (how mass tells space-time to curve). So it is quite
 remarkable that a covariant formulation of the strong equivalence
 principle (SEP) applied to generalized Brans-Dicke gravity models
 singles out one non-linear tensor theory ($R_{\sigma\nu}=0$) and 
 one linear scalar theory ($W^{\sigma\lambda\mu\nu} = 0, R = 0$). They may be considered
 as the analogs of the non-abelian QCD and abelian QED gauge theories,
 respectively: gluons carry colours ($D_{\mu} F^{\mu\nu} =
0$) but photons do not carry electric charge ($\partial_{\mu}
F^{\mu\nu}=0$). From a
 theoretical point of view, the abelian alternative to general
 relativity should not come as a surprise.  The WEP alone
 implies that all non-gravitational fields couple in the same
 way to gravity. The strong version of the equivalence
 principle simply extends this universality of coupling to the
 gravitational field itself. 
However the photon, like any massless particle, does not couple to a conformally flat metric field. Consequently, the survival of a scalar theory complying with the SEP is obvious at the theoretical level, though ruled out at the phenomenological one. 

\bigskip

 If the SEP proves to be fundamental, then Eq.(6) itself should  arise from a
 new theory of gravity, in a way similar to Eq.(3). The latter is derived from the pure Yang-Mills action functional which  is an integral over the square of the curvature,
$F^{\mu\nu}F_{\mu\nu}$. Here, no attempt has been made to obtain the former  
from a variational principle. We simply note that if
one considers, again by analogy, the conformally invariant Weyl theory in regions devoid of any matter : 
\begin{equation}
S_{\scriptsize \textrm{Weyl}} = \xi \int d^4x \sqrt{-g} \{W^{\sigma \lambda
\mu \nu }W_{\sigma \lambda \mu \nu }\},
\end{equation}
the resulting fourth-order differential equations for the metric \cite{25}:
\begin{equation}
2D_{\nu}D_{\sigma}W^{\sigma \lambda \mu \nu } + W^{\sigma \lambda \mu \nu } 
R_{\sigma \nu } = 0
\end{equation}
contain non-trivial solutions \cite{26} but reduce to the same alternative
as for the tensor-scalar models once the constraint (14) are imposed
by hand.

  In the presence of matter fields, any strict analogy with the
  Yang-Mills equation of motion (3) has to break down at some point
  since the affine connection is itself constructed from the first
  derivatives of the metric tensor, while the gauge fields are not
  expressed in terms of more fundamental fields. However, an elegant
  formulation of gravity resembling the Yang-Mills theory has been
  advocated in \cite{27,28}. In this purely geometric formulation, the
  action is independent of the metric : the
  Lagrangian is exclusively quadratic in the field strength of an
  appropriate gauge group $G$ having the Lorentz group as a
  subgroup. Here, the metric is no more a fundamental field than a
  hadron field is a fundamental field in QCD [29]. Indeed, four of the
  gravitational gauge potentials which belong to the adjoint
  representation of $G$ are subsequently identified with the
  vierbein. The original action reduces then to the Einstein-Hilbert
  one with the addition of a cosmological term, in agreement with
  our Eq.(13). Thus, the SEP might indeed be a useful guiding gauge
  principle to single out such a higher-order action functional for
  gravitational interactions.

\bigskip

On the other hand, if the SEP turns out to be only approximate, the covariant
condition (6) suggests a simple and natural way to implement a minimal violation
of it within a sub-class of tensor-scalar theories characterized by an attractor
mechanism toward general relativity. 
In particular, Eq.(33) nicely illustrates how matter is allowed to change Eq.(6).

\section*{Acknowledgements}
 This work was supported by the Belgian Federal Office for Scientific, Technical
 and Cultural Affairs through the Interuniversity Attraction Pole P5/27.

\end{document}